\documentstyle[twocolumn,seceq,epsf]{jpsj}

\def\runtitle{Stripe orders in the extended Hubbard model}
\def\runauthor{Taira {\sc Kato} and Masaru {\sc Kato}}

\title
{
Stripe orders in the extended Hubbard model
}

\author
{ 
Taira {\sc Kato}
and Masaru {\sc Kato}\footnote{E-mail:kato@ms.osakafu-u.ac.jp}
}

\inst
{
Department of Mathematical Sciences,
Osaka Prefecture University, Sakai, Osaka 599-8531, JAPAN
}

\recdate
{
\today
}

\abst
{
We study stripe orders of charge and spin density waves 
in the extended Hubbard model with the nearest-neighbor Coulomb repulsion $V$
 within the mean field approximation.
We obtain $V$ vs. $T$(temperature) phase diagram for the on-site Coulomb 
interaction $U/t=8.0$ and the filling $n=0.8$, here $t$ is
a nearest-neighbor transfer energy.
Our result shows that the diagonal stripe spin density wave state (SDW)
is stable for small $V$, but for large $V$
the most stable state changes to 
a charge density wave-antiferromagnetic (CDW-AF) state. 
Especially we find at low temperature and for a certain range of value of $V$,
a vertical stripe CDW-AF state becomes stable.
}

\kword
{
stripe order, incommensurate spin density wave, extended Hubbard model
}

\begin{document}
\sloppy
\maketitle

\section{Introduction}
Much attention has been focused on the phenomena of charge and spin orderings
in transition metal oxides
including high temperature superconductiong cuprates
since the discovery of the $1/8$ problem of 
La$_{2-x}$Ba$_x$CuO$_4$.\cite{moodenbaugh:1}.
La$_{2-x}$Sr$_x$NiO$_4$\cite{hayden:1,sachan:1, tranq:3,
yoshizawa:1,chen:1, katsufuji:1, ramirez:1, vigliante:1, blumberg:1,
yamamoto:1, yoshinari:1, hess:1, katsufuji:2, yamanouchi:1} and
La$_2$NiO$_{4+\delta}$ \cite{tranq:1,tranq:1-1,tranq:1-2,tranq:2,tranq:2-1,
tranq:4,tranq:4-1} is the typical material 
of this phenomena.
La$_{2-x}$Sr$_x$NiO$_4$ shows the stripe order of the spin and
the charge and is insulating while $x<0.5$.\cite{hayden:1}
The direction of the stripe is diagonal to the crystal axis.\cite{hayden:1}
The relation between the hole concentration $n_h$ and
the incommensurability $\epsilon$ is approximately linear.\cite{yoshizawa:1}
The stripe order consists of the charge stripe and
incommensurate spin density wave(SDW).
The critical temperature of the charge stripe is higher than that of
the incommensurate SDW.\cite{yoshizawa:1}

In early theoretical mean-field study of the Hubbard model
\cite{schulz, schulz-1,poiblanc,kato:1,inui} and the d-p model,\cite{zaanen}
the stable states are
the insulating stripe or incommensurate spin density wave (SDW) states.
In the hole density vs.~on-site Coulomb interaction phase diagram
at zero temperature,
the diagonal stripe SDW state becomes stable for large on-site Coulomb
intercation and small hole doping.\cite{kato:1}
Temperature dependence of the stable states is studied within the mean-field
approximation in Ref.~\citen{kato:2}.
The result shows that as decreasing temperature,
the stable state changes from the paramagnetic state to N\'{e}el state,
the spiral SDW state,
the stripe SDW state and the undulated stripe SDW state, successively.
Especially, the charge and the spin stripe orders appear simultaneously.
But this result is inconsistent with the experimental fact.

Also such charged stripe may become unstable
by the long range Coulomb interaction, although there is the distortion of 
the ionic lattice.
Therefore we study the extended Hubbard model with nearest-neighbor Coulomb
interaction in order to consider the stability of the stripe state against
the long range Coulomb interaction.
We also consider the temperature dependence of the stable state
within the mean-field approximation
in order to search for the possibility of the charged stripe
without the static spin ordering at finite temperature.

This paper is arranged as follows.
In section 2, we show our model and our method of
the mean-field approximation.
In section 3, we show our numerical result, especially the phase diagram
and we examine the stable state using the band calculation.
Section 4 is devoted to the discussion and conclusion.

\section{Model}
We use following the extended Hubbard model,
\begin{eqnarray}
  {\mathcal H}&=&-t\sum_{\left<i,j\right>}
  \left(c_{i\sigma}^\dagger c_{j\sigma}
  + H.c.\right)+
  U\sum_i\hat{n}_{i\uparrow}\hat{n}_{i\uparrow} \nonumber\\
  &&+V\sum_{\left<i,j\right>}\hat{n}_i\hat{n}_j,
\end{eqnarray}
where $\left<i,j\right>$ are nearest neighbor sites,
$c_{i\sigma}^\dagger$ ($c_{i\sigma}$) 
is an electron creation (annihilation) operator at site $i$ with spin $\sigma$,
$\hat{n}_{i\sigma}=c_{i\sigma}^\dagger c_{i\sigma}$
($\sigma=\uparrow, \downarrow$) is an  electron number operator 
at site $i$ with spin $\sigma$ and $\hat{n}=\hat{n}_\uparrow +\hat{n}_\downarrow$.
The nearest-neighbor transfer energy is $t$ and the on-site and
the nearest-neighbor Coulomb interactions are $U$ and $V$, respectively.

We take following mean fields for each site;
\begin{eqnarray}
\rho_{i\sigma}=\left<\hat{n}_{i\sigma}\right>,\\
X_i=\left<c^\dagger_{i\uparrow}c_{i\downarrow}\right>,
\end {eqnarray} 
where $\left<\ \right>$ means thermal average.
Also we take following mean fields for each nearest-neighbor bond
$\left<i,j\right>$;
\begin{eqnarray}
W_{i\sigma j\sigma'}=\left< c^\dagger_{i\sigma}c_{j\sigma'}\right>.
\end{eqnarray}
Then our mean-field Hamiltonian becomes as follows,
\begin{eqnarray}
\mathcal{H}_{\mbox{$\scriptstyle MF$}}&=&
-t\sum_{<i,j>}\left(c_{i\sigma}^\dagger c_{j\sigma}+ H.c.\right)+
\sum_{i,\sigma}U\rho_{i\sigma}c^\dagger_{i,-\sigma}c_{i,-\sigma}\nonumber\\
&-&\sum_{i}U\left(X_ic^\dagger_{i\downarrow}c_{i\uparrow}  +
X_i^*c^\dagger_{i\uparrow}c_{i\downarrow}  \right)\nonumber \\
&+&\sum_{<i,j>,\sigma}V\rho_j\hat{n}_{i\sigma}
-\sum_{<i,j>,\sigma\sigma'}VW_{i\sigma j\sigma'}
c^\dagger_{j\sigma'}c_{i\sigma}\nonumber\\
&-&\sum_{i}U\rho_{i\uparrow}\rho_{i\downarrow}
+\sum_{i}U\left|X_i\right|^2
-\sum_{<i,j>}V\rho_j\rho_i \nonumber \\
&+& \sum_{<i,j>,\sigma\sigma'}V\left|W_{ij,\sigma\sigma'}\right|^2,
\end{eqnarray}
where 
\begin{equation}
\rho_i=\rho_{i\uparrow}+\rho_{i\downarrow}.
\end{equation}
We write $n$-th eigenstate of this hamiltonian as,
\begin{equation}
\left|\Psi_n\right> = \sum_{i\sigma}\psi_{i\sigma}^nc^\dagger_{i\sigma}
\left|0\right>,
\end{equation}
with the eigenvalue $E_n$.
Then Schr\"{o}dinger equation becomes,
\begin{eqnarray}
-t\sum_j^{\mathrm{n.n.}\mathit \left( i\right)}\psi_{j\sigma}^n+
U\rho_{i-\sigma}\psi_{i\sigma}^n
-UX_i^{*^\sigma}\psi_{i-\sigma}^n\nonumber\\
+V\sum_j^{\mathrm{n.n.}\mathit \left( i\right)}\rho_j\psi_{i\sigma}^n
-V\sum_{j\sigma'}^{\mathrm{n.n.}\mathit \left( i\right)}
W_{j\sigma'i \sigma }\psi_{j\sigma'}^n
=E_n\psi_{i\sigma}^n,\label{eigen}
\end{eqnarray}
where n.n.($i$) means nearest neighbor sites of $i$-th site and 
$X_i^{*^\sigma}$ means $X_i^{*}$($X_i$) when $\sigma$ is 
$\uparrow$($\downarrow$).
We can rewrite mean fields using eigenstates $\left\{\psi_{i\sigma}^n\right\}$
as follows;
\begin{eqnarray}
\rho_{i\sigma}=\sum_n\left|\psi_{i\sigma}^n\right|^2 f\left(E_n-\mu\right), \label{sc:1}\\
X_i=\sum_n\psi_{i\uparrow}^{n*}\psi_{i\downarrow}^n
f\left(E_n-\mu\right),\label{sc:2}\\
W_{i\sigma j\sigma'}=\sum_n \psi_{i\sigma}^{n*}\psi_{j\sigma'}^n
f\left(E_n-\mu\right)\label{sc:3},
\end{eqnarray}
where $f$ is the fermi distribution function and $\mu$ is a chemical potential.
Also we impose the electron number conservation,
\begin{equation}
N_e=\sum_nf\left(E_n-\mu\right),\label{number}
\end{equation}
where $N_e$ is the total number of electrons.

We solve eq.~(\ref{eigen}),
then find out the chemical potential $\mu$ by eq.~(\ref{number})
and substitute the results to self-consistent equations,
eqs.~(\ref{sc:1}), (\ref{sc:2}) and (\ref{sc:3}).
We continue this procedure until
all of the mean fields converge.
For each values of $U$, $V$ and temperature $T$, we use several initial states and
obtain several self-consistent solutions.
Then we calculate the free energy of each state using following formula,
\begin{eqnarray}
F&=&\mu N_e -k_BT\sum_n\ln \left[1+e^{-\beta\left(E_n-\mu\right)}\right]
-U\sum_i\rho_{i\uparrow}\rho_{i\downarrow}\nonumber \\
&+&U\sum_i\left|X_i\right|^2
-V\sum_{\left<ij\right>}^{n.n.}\left[\rho_i\rho_j
-\sum_{\sigma\sigma'}\left|W_{i\sigma j\sigma'}\right|^2\right].
\end{eqnarray}
Comparing free energies we determine the most stable configuration
for each values of parameters.

\section{Results}
We use a square lattice with $10\times 10$ sites and impose the periodic
boundary condition.
The total electron number is fixed to 80,
and this means that the filling is 0.8.
Also we fix the on-site Coulomb interaction $U$ to 8,
where we choose $t=1$ as an energy unit.
Form the result of Ref.~\citen{kato:1},
the diagonal stripe state is stable at zero temperature and $V=0$.
We vary $V$ from 0 to 3 in units of 0.25 and $T$ from 0.1 to 3.75.
For each values of both of parameters $V$ and $T$,
we choose several initial states for the self-consistent procedure,
and obtain several self-consistent solutions, which may
be the global minimum or
the local minimum of the free energy.

\subsection{Phase diagram}
In the following, we show the transition of most stable state
as we vary the temperature for each value of $V$.

\begin{figure}
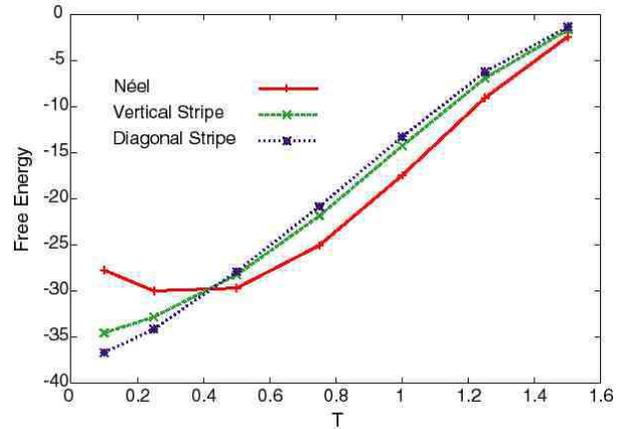

\epsfxsize=8cm
\epsfigure{fig1.EPSF}
\caption{Temperature dependence of free energies for several stable states
when $V=0$.
Each free energy is measured from the free energy of paramagnetic state.}
\label{fe:v0}
\end{figure}

\begin{figure}
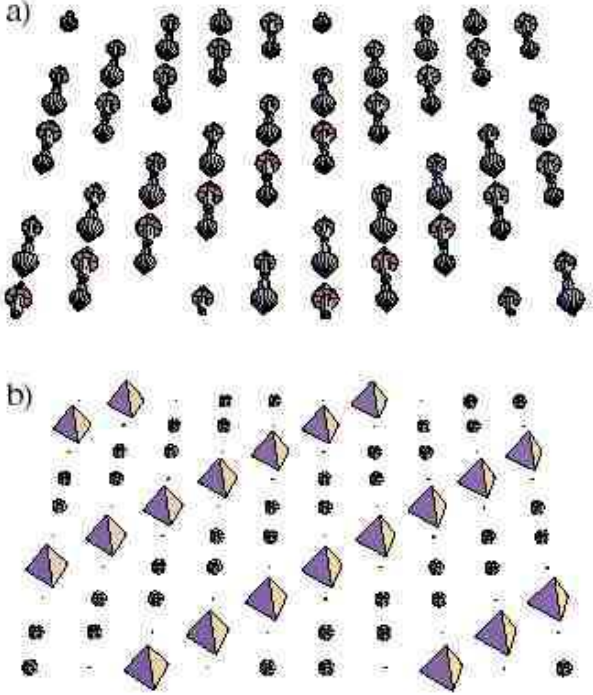

\epsfxsize=8cm
\epsfigure{fig2.EPSF}
\caption{Diagonal stripe state at $V=0$ and $T=0.1$.
a) Spin density. b) Charge density.
The size (direction) of the arrow shows the magnitude (direction) of the moment. 
The sphere (tetrahedron) means the charge density is larger (smaller)
than the average value of the charge density (0.8), respectively,
and the size of the sphere (tetrahedron) shows
the deviation from the average value.}
\label{diagonal:v0:1}
\end{figure}

\begin{figure}
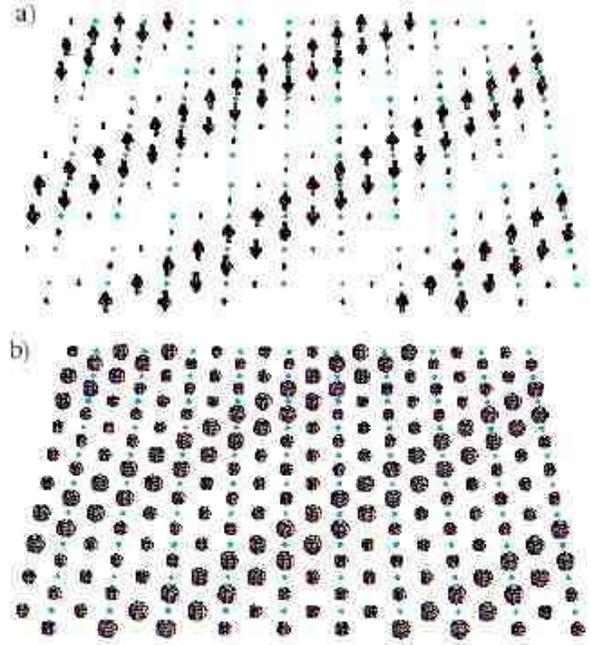

\epsfxsize=8cm
\epsfigure{fig3.EPSF}
\caption{Diagonal stripe state at $V=0$ and $T=0.1$.
a) Bond spin density. 
b) Bond charge density.
The meaning of the arrow is same as that of Fig.~\ref{diagonal:v0:1}.
The size of the sphere shows the magnitude of the bond spin density.}
\label{diagonal:v0:2}
\end{figure}

When nearest neighbor Coulomb interaction $V$=0, 
decreasing temperature $T$ the most stable state changes
from the paramagnetic state to the N\'{e}el state,
then changes to a diagonal stripe SDW state.
This result almost agrees with the previous result with $U=5$\cite{kato:2},
although in the previous result shows that
the lowest temperature phase becomes undulated stripe SDW state
which becomes stable at $T<0.1$, the vertical stripe state is stable
instead of the diagonal stripe state and the spiral SDW state becomes
stable in a small temperature region between the N\'{e}el state and
the diagonal stripe state.
In Fig.~\ref{fe:v0}, we show temperature dependence of free energies
of several phases at $V=0$. In this figure, we plot free energy difference
from the paramagnetic state.
From this figure we find that,
the diagonal stripe state becomes more stable than
the vertical stripe state when temperature is lower than a critical point
where these stripe states become more stable than the N\'{e}el state,
although the free energy of the vertical stripe state is lower than
that of the diagonal stripe state in the region
where the N\'{e}el state is most stable.
We show the spin density, the charge density of the diagonal stripe state
at $T=0.1$ and $V=0.0$ in Fig.~\ref{diagonal:v0:1}, the bond spin
and the bond charge densities in Fig.~\ref{diagonal:v0:2}.
Here the charge density is $\rho_i$ and the spin density
is defined as follows,
\begin{equation}
{\mib S}_i=(\Re X_i, \Im X_i,
\frac{\rho_{i\uparrow}-\rho_{i\downarrow}}{2}),
\end{equation}
where $\Re$ and $\Im$ mean the real and the imaginary part
of a complex number, respectively.
We define a complex bond charge density
for a bond between nearest neighbor sites $\left<ij\right>$ as,
\begin{equation}
\rho_{ij}=\sum_{\sigma}W_{i\sigma j\sigma}.
\end{equation}
Also we define the complex bond spin density as,
\begin{eqnarray}
&&{\mib S}_{ij}=\nonumber\\
&&(\frac{W_{i\uparrow j\downarrow}+
W_{i\downarrow j\uparrow}}{2},
\frac{W_{i\uparrow j\downarrow}-W_{i\downarrow j\uparrow}}{2i},
\frac{W_{i\uparrow j\uparrow}-W_{i\downarrow j\downarrow}}{2}).
\end{eqnarray} 
Note that $\psi_{i\sigma}$ is real, therefore $\rho_{ij}$, $S_{ij}^x$ and 
$S_{ij}^z$ are real and $S_{ij}^y$ is pure imaginary.
All of stable states that we have obtained do not have the
imaginary  $S_{ij}^y$ component.
So we plot real part of $\rho_{ij}$ and real part of ${\bf S}_{ij}$ 
at the center of the bond $\left<ij\right>$ in Fig.~\ref{diagonal:v0:2}..
\begin{figure}
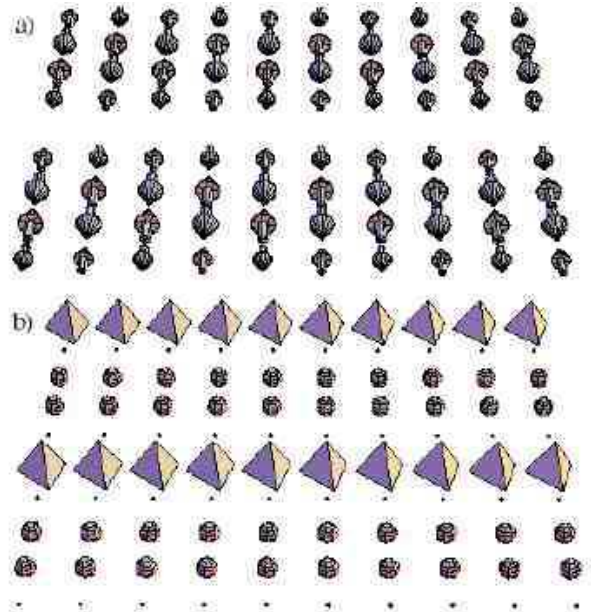

\epsfxsize=8cm
\epsfigure{fig4.EPSF}
\caption{Vertical stripe state at $V=0$ and $T=0.1$.
a) Spin density. b) Charge density.
The meaning of symbols is same as that of Fig.~\ref{diagonal:v0:1}.}
\label{vertical:v0:1}
\end{figure}
\begin{figure}
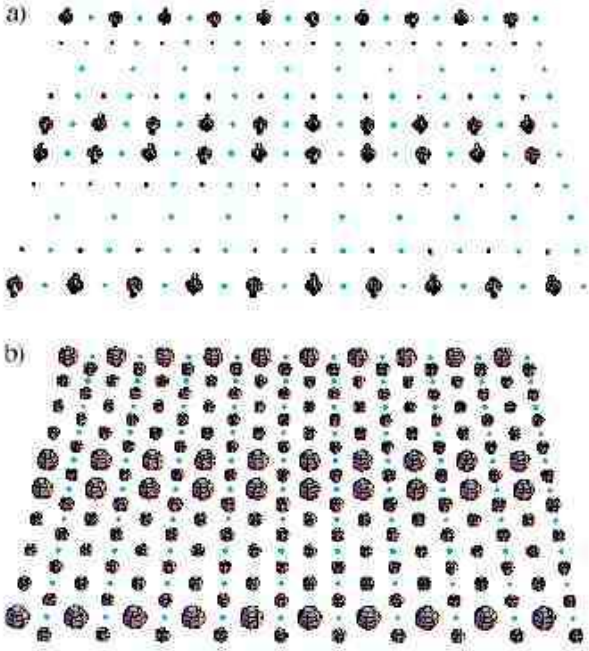

\epsfxsize=8cm
\epsfigure{fig5.EPSF}
\caption{Vertical stripe state at $V=0$ and $T=0.1$.
a) Bond spin density. b) Bond charge density.
The meaning of symbols is same as that of Fig.~\ref{diagonal:v0:2}.}
\label{vertical:v0:2}
\end{figure}

For the paramagnetic and the N\'{e}el states
the bond charge density distribution is uniform
and the bond spin densities vanish.
In contrast to this, both of the bond spin
and bond charge density distributions of the diagonal stripe state
have a diagonal stripe structure.
These densities are large at the bonds
that connect the site in a stripe and the site 
in an antiferromagnetic (AF) domain.
We also show the vertical stripe states in Fig.~\ref{vertical:v0:1} and 
Fig.~\ref{vertical:v0:2}. 
Similar to the diagonal stripe state,
the bond spin and bond charge densities
have stripe structure and are large
at the boundary between of the stripe and 
the AF domain.
 
For finite but small $V\left(<1.0\right)$,
the temperature dependence of the most stable state
is same as $V=0$ case.
When $V$ is increased,
the structure of the diagonal stripe state is invariable,
though the magnitudes of all of its densities become small.
This means the diagonal stripe state is robust about small variation of
the nearest neighbor repulsion $V$. 
Also the structure of the paramagnetic and the N\'e{e}l states are invariant
with respect to the variation of $V$.

\begin{figure}
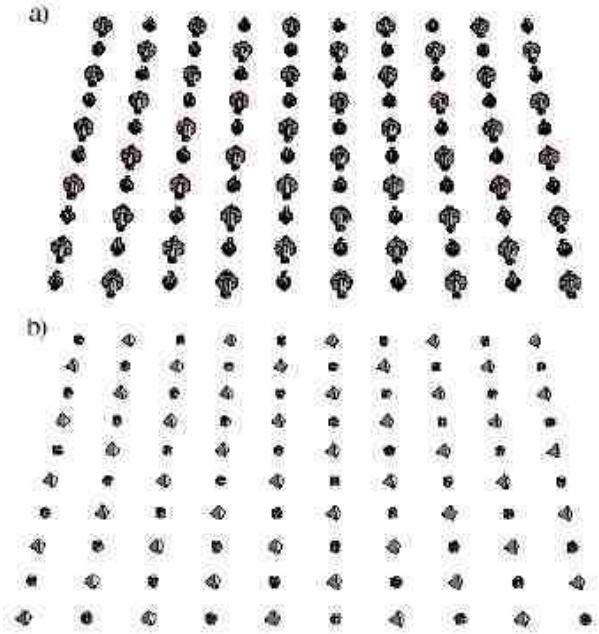

\epsfxsize=8cm
\epsfigure{fig6.EPSF}
\caption{CDW-AF state at $V=1.0$ and $T=0.5$.
a) Spin density. b) Charge density.
The meaning of symbols is same as that of Fig.~\ref{diagonal:v0:1}.}
\label{CDW-AF:v1:1}
\end{figure}

\begin{figure}
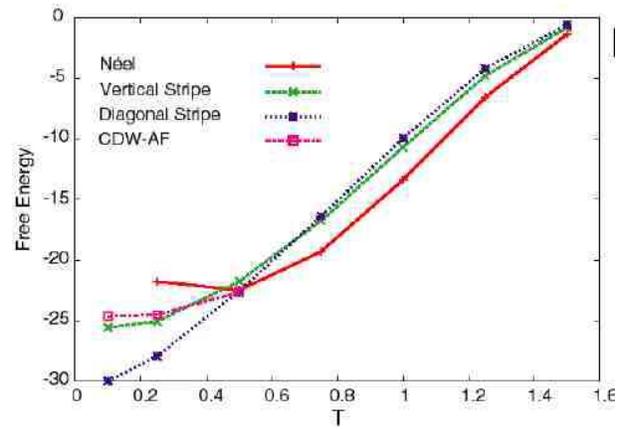

\epsfxsize=8cm
\epsfigure{fig7.EPSF}
\caption{Temperature dependence of free energies for several stable states
when $V=1$.
Free energies are measured from the free energy of paramagnetic state.}
\label{fe:v1}
\end{figure}

Further increasing $V$ up to $1.0$,
then a charge density wave-antiferromagnetic state (CDW-AF)
becomes stable in a small temperature region which
locates between the N\'{e}el state and the diagonal stripe state.
We show the structure of the CDW-AF state 
in Fig.~\ref{CDW-AF:v1:1}.
Also in Fig.~\ref{fe:v1},
we show the variation of free energies of several states with temperature.
The charge density distribution of the CDW-AF state is same as
the ordinary charge density wave state (CDW).
However, the spin density does not vanish and 
is similar to that of the ferrimagnetic state; large moment appears 
at the site where the charge density is large
and opposite small moment appears
at the site where the charge density is small.
The bond spin and bond charge densities are similar to the paramagnetic and
the N\'{e}el states; the bond charge density is uniform and
the bond spin density is very small.

\begin{figure}
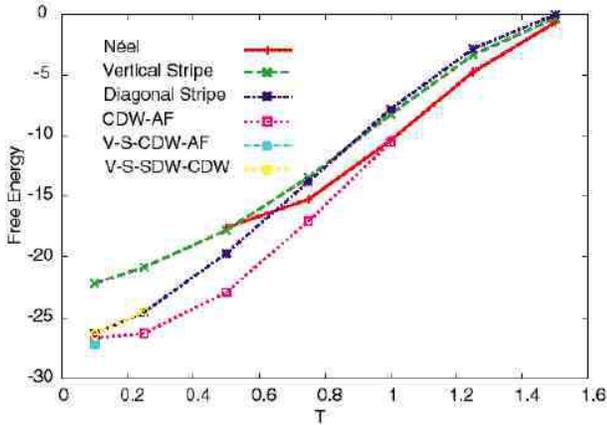

\epsfxsize=8cm
\epsfigure{fig8.EPSF}
\caption{Temperature dependence of free energies for several stable states 
when $V=1.75$.
Free energies are measured from the free energy of paramagnetic state.
V-S-CDW-AF stands for vertical stripe CDW-AF
and V-S-SDW-CDW stands for vertical stripe SDW-CDW.}
\label{fe:v1.75}
\end{figure}

\begin{figure}
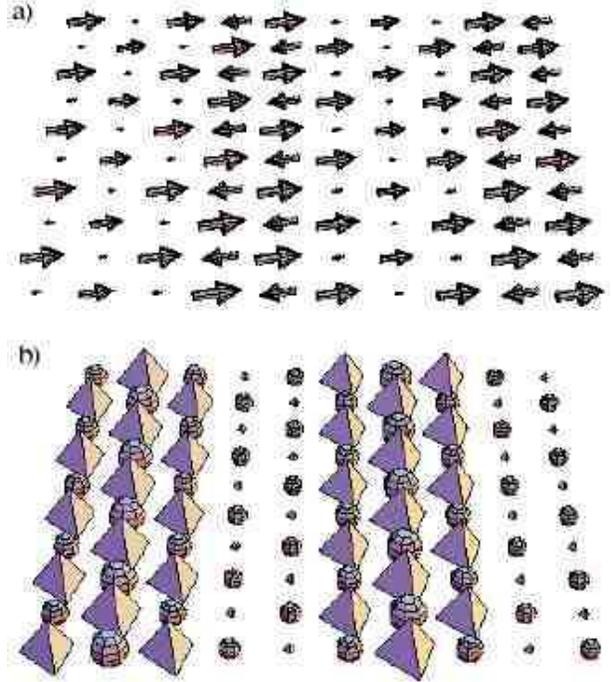

\epsfxsize=8cm
\epsfigure{fig9.EPSF}
\caption{Vertical stripe CDW-AF state at $V=1.75$ and $T=0.1$.
a) Spin density. b) Charge density.
The meaning of symbols is same as that of Fig.~\ref{diagonal:v0:1}.}
\label{ver-CDW-AF:v1.75:1}
\end{figure}

Further increasing $V$ up to 1.5,
the temperature dependence of the most stable
state is almost same, but the temperature region where CDW-AF state is most
stable, becomes large. 
When $V=1.75$, a vertical stripe CDW-AF state becomes more stable than the
diagonal stripe state at low temperature ($T=0.1$).
Although, at $T=0.1$ we obtain a vertical stripe SDW-CDW state
which is a mixture of the vertical stripe SDW state and the CDW state,
its free energy is slightly higher than
that of the vertical stripe CDW-AF state.
Also we have not obtained the diagonal type of stripe CDW-AF
or stripe SDW-CDW state.
In Fig.~\ref{fe:v1.75}, we show temperature dependence of free energies
of several states at $V=1.75$. 
Also we show the spin and charge distributions of this new phase,
the vertical stripe CDW-AF state, at $V=1.75$ and $T=0.1$
in Fig.~\ref{ver-CDW-AF:v1.75:1}, 
the bond spin and the bond charge distributions in
Fig.~\ref{ver-CDW-AF:v1.75:2}.
\begin{figure}
\epsfxsize=8cm
\epsfigure{fig10.EPSF}
\caption{Vertical stripe CDW-AF state at $V=1.75$ and $T=0.1$.
a) Bond spin density. 
b) Bond charge density.
The meaning ofsymbols is same as that of Fig.~\ref{diagonal:v0:2}.}
\label{ver-CDW-AF:v1.75:2}
\end{figure}
From these figures,
we can consider this phase as a vertical striped previous CDW-AF state.
From the point of view of the CDW order, 
this phase has vertical domains where the CDW order is large,
which are separated by the stripes where the CDW order is small.
However, from the point of view of the antiferromagnetic order,
this phase has narrow domains of the ferrimagnetic phase,
which is separated by the broad stripes of a half ferromagnetic phase,
where half of the sites have ferromagnetic order and the others
have no spins.
Similar to the stripe SDW state, the bond spin density is large
at the bond between the ferrimagnetic domain
and the half-ferromagnetic stripe.
The bond charge density is large at the bond
where bond spin density is large.
However the bond at the both side of the site
where magnetic moment is large in the half-ferromagnetic stripe,
also has large bond charge density.

In the parameter region $V=1.75\mbox{ and }2.0$, 
the temperature dependence of the stable state is almost same,
but the temperature region where the vertical stripe CDW-AF state is stable
becomes large at $V=2.0$.

\begin{figure}
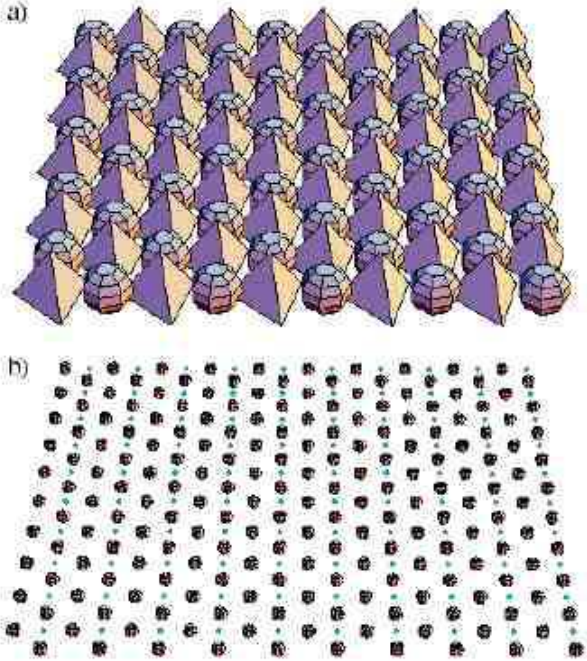

\epsfxsize=8cm
\epsfigure{fig11.EPSF}
\caption{CDW state at $T=1.5$ and $V=2.25$.
a) Charge density. b) Bond charge density.
The meaning of symbols is same as that of Figs.~\ref{diagonal:v0:1}
and \ref{diagonal:v0:2}.}
\label{CDW:v2.25}
\end{figure}

\begin{figure}
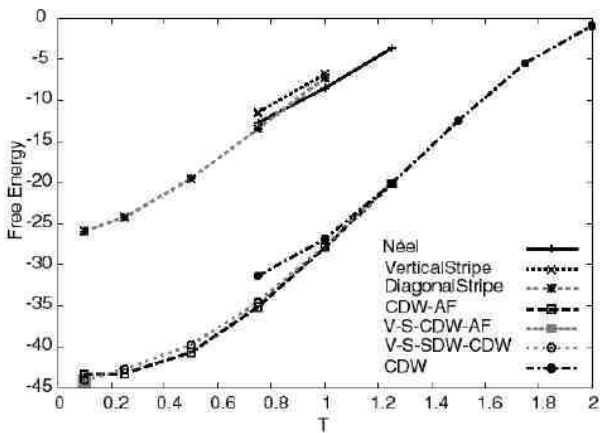

\epsfxsize=8cm
\epsfigure{fig12.EPSF}
\caption{Temperature dependence of free energies for several stable states 
when $V=2.25$.
Free energies are measured from the free energy of paramagnetic state.}
\label{fe:v2.25}
\end{figure}

At $V=2.25$, the higher temperature phase changes from the N\'{e}el state to
the CDW state.
The spin and the bond spin densities vanish and as shown in Fig.~\ref{CDW:v2.25},
the bond charge distribution is uniform.
This behavior can be expected, because the large nearest-neighbor Coulomb
repulsion causes the alternating charge density distribution.
The temperature dependence of free energies of various states
is shown in Fig.~\ref{fe:v2.25}.

\begin{figure}
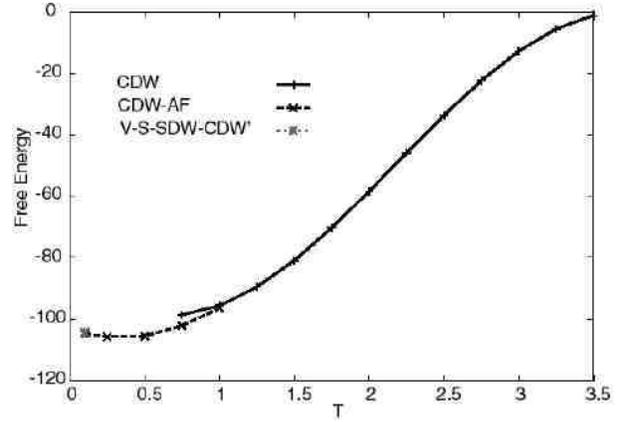

\epsfxsize=8cm
\epsfigure{fig13.EPSF}
\caption{Temperature dependence of free energies for several stable states 
when $V=3.0$.
Free energies are measured from the free energy of paramagnetic state.}
\label{fe:v3.0}
\end{figure}

At $V=3.0$, stripe phases are no longer stable
and the most stable state changes
from the CDW to the CDW-AF state as decreasing temperature.
The temperature dependence of the free energies of various states
is shown in Fig.~\ref{fe:v3.0}.
Free energies of stripe orders are much higher than that of
CDW based states.

\begin{figure}
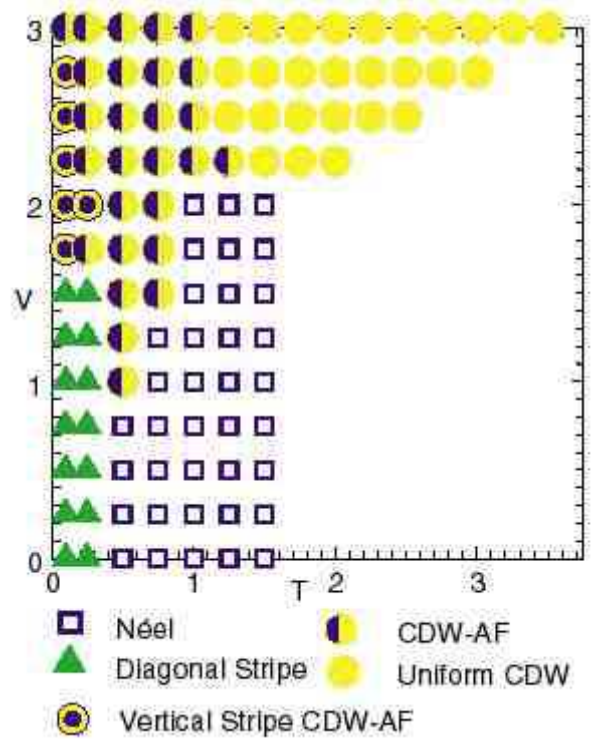

\epsfxsize=8cm
\epsfigure{fig14.EPSF}
\caption{Phase diagram in $T$ vs.\ $V$ plane.}
\label{PD}
\end{figure}

The total phase diagram the temperature $T$ vs.\
the nearest-neighbor repulsion $V$ is shown in Fig.~\ref{PD}.

\subsection{Energy band}
\begin{figure}
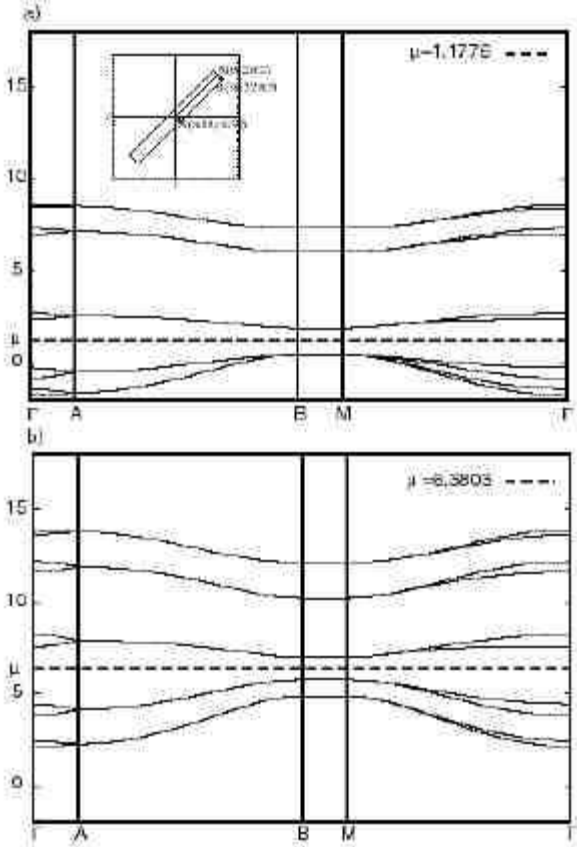

\epsfxsize=8cm
\epsfigure{fig15.EPSF}
\caption{Energy band of the diagonal stripe state at in $T=0.1$ and
$V=0.0$(a) and $1.5$(b). The inset is the first Brillouin zone.
$\mu$ is the chemical potential.}
\label{EB:d-sdw}
\end{figure}

Here we discuss the electronic structure of several stable states.
In Fig.~\ref{EB:d-sdw}, we show the energy band of
the diagonal stripe states at $T=0.1$ and $V=0.0 \mbox{ and }1.5$.
There are three types of the energy band.
One is the occupied valence band, second one is the midgap band and
third one is the empty conduction band.
The wave function of the mid-gap state is localized at stripes.
The site on the diagonal stripe does not have the nearest neighbor site
on the stripe.
Therefore the band is almost flat.
The nearest neighbor Coulomb interaction $V$ does not much affect
the band structure, only the lower band become split
around B-point($k=\left(3\pi/5,2\pi/5\right)$).
At $V=0.0$ and at $\Gamma$-point($k=\left(0,0\right)$),
the wave functions of the lower band of the valence bands
have large amplitude on the edge sites of AF domain, but 
the wave functions of the higher band of the valence bands
have large amplitude on the inner sites of AF domain.
So the lower band has a large band width because of the transfer energy,
and at M-point($k=\left(\pi/2,\pi/2\right)$) its energy become higher.
But at $V=1.5$, the wave function which has large amplitudes on the inner sites
have larger energy because of the direct nearest-neighbor repulsion.
Therefore for larger $V$, the energy difference of these band becomes large.
Also the same reason that the midgap state becomes flat, causes
the band invariant under the variation of $V$.

\begin{figure}
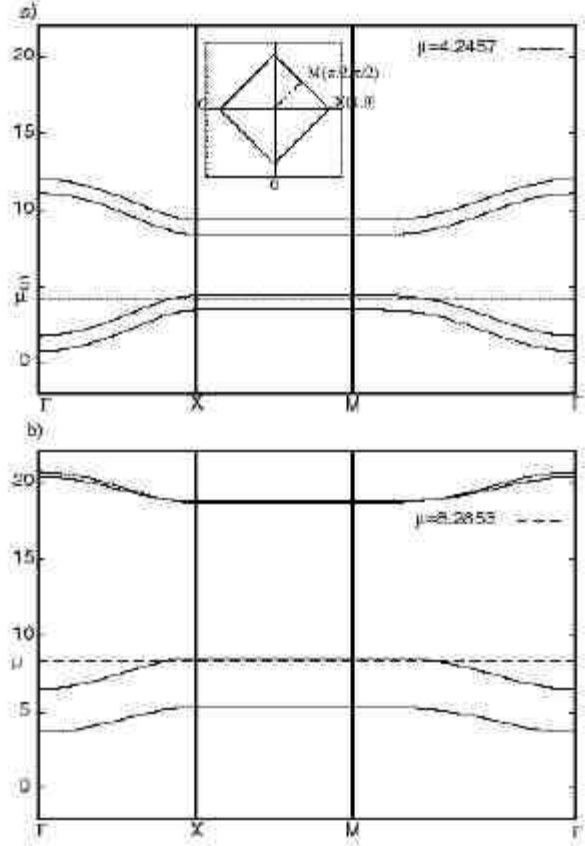

\epsfxsize=8cm
\epsfigure{fig16.EPSF}
\caption{Energy band of the CDW-AF state at in $T=0.5$ and
$V=1.0$(a) and $3.0$(b).
The inset is the first Brillouin zone.
$\mu$ is the chemical potential.}
\label{EB:cdw-af}
\end{figure}

We show the energy band of the CDW-AF state
at $T=0.5$ and $V=1.0 \mbox{ and } 3.0$ in Fig.~\ref{EB:cdw-af}.
There are two kinds of the bands, almost occupied valence bands
and empty conduction band.
Both kinds of the bands have two separated band.
At $V=1.0$ the wave function of the higher energy one of valence bands
has large amplitude on the site where large spin exists,
with the same spin.
And that of the lower energy one has 
large amplitude on the site where a small spin exists, with the same spin.
Therefore the energy difference comes from the direct nearest neighbor
Coulomb interaction.
For conduction bands, also the energy difference comes from 
the direct nearest neighbor Coulomb interaction.
The band gap between conduction bands and valence bands comes from
the on-site Coulomb interaction.
For $V=3.0$, the split of two of valence bands becomes large.
The wave functions of the lower band of the valence bands have large
amplitude on the site where large spin exists, with the same spin,
though that of the upper band have large amplitude on the same site
but with the opposite spin.
Therefore the energy difference of these two bands comes from the
on site Coulomb interaction.
The wave functions of both of the conduction bands have large amplitude
at the site where the charge density almost vanishes.
Therefore the band gap energy comes 
from the nearest neighbor Coulomb interaction,
and the on-site Coulomb interaction does not affect much on the
energies of these two conduction bands.
As we have shown, the band structure of the CDW-AF state
varies strongly with the nearest neighbor Coulomb interaction.

\begin{figure}
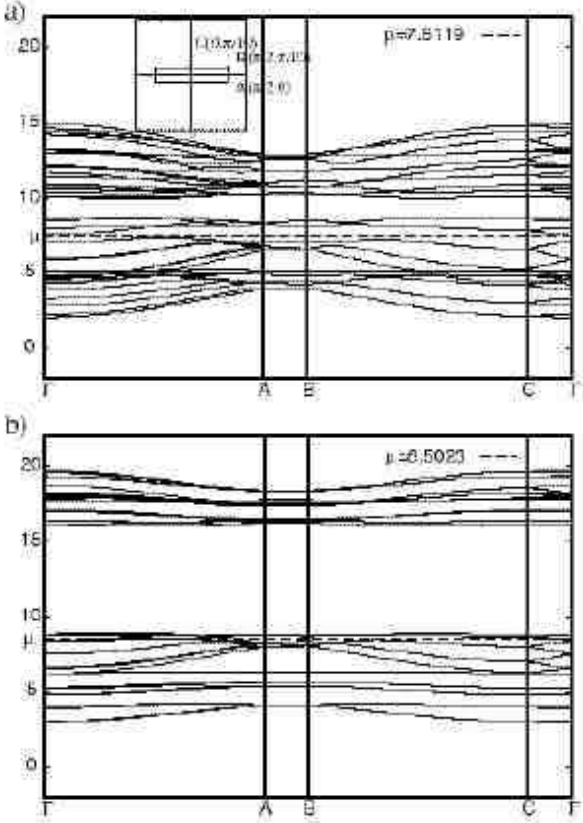

\epsfxsize=8cm
\epsfigure{fig17.EPSF}
\caption{Energy band of the vertical CDW-AF state at in $T=0.1$ and
a)$V=1.75$ and b)$V=2.75$.
The inset is the first Brillouin zone.
$\mu$ is the chemical potential.}
\label{EB:vs-cdw-af}
\end{figure}

We show the energy band of the vertical stripe CDW-AF state
at $T=0.1$ and $V=1.75 \mbox{ and } 2,75$ in Fig.~\ref{EB:vs-cdw-af}.
At $V=1.75$, there are midgap bands with valence bands and conduction
bands.
The wave function of valence bands has large amplitude
with the same spin as the spin density of the site,
independent of the magnitude of the spin density.
The wave function of the conduction bands is opposite.
But for midgap bands, the wave fuction has large amplitude
at the site where the spin density is small with the same spin.
The band gap energy of valence bands and conduction bands comes
from the on-site Coulomb interaction, similar to the CDW-AF state
at $V=1.0$.

But at $V=2.75$, the midgap bands almost merge to valence bands
and this phase becomes metallic unlike to other stripe phases.
Similar to the CDW-AF states, the wave function of the valence bands has
large amplitude at the site where the charge density is large,
and that of the conduction band has large amplitude at the site where
the charge density is small.
So, the band gap energy comes from the nearest neighbor Coulomb interaction.
This vertical stripe CDW-AF state changes largely the band structure
when the value of $V$ changes just 1.
This feature is different from the diagonal stripe state of which
band structure is invariant to the variation of $V$.

\section{Conclusion}
In this paper, we have solved self-consistent equation
of the extended Hubbard model on the square lattice.
We have obtained the nearest neighbor Coulomb repulsion $V$
vs.~the temperature $T$ phase diagram.
At low temperature there are stripe phases and the diagonal
stripe state is stable for the finite but small $V$.
For larger $V$, the vertical stripe CDW-AF state becomes stable.
The band structure of the diagonal stripe state is invariant against
$V$, but the band structure of the vertical stripe CDW-AF
state changes largely by small variation of $V$.

In relation to the real materials, especially nickelates,
the stripe charge ordered state is not stable in this parameter region
within the mean-field approximation.
The nearest neighbor Coulomb interaction $V$ only stabilizes
the CDW based states.
In order to obtain the stripe charge ordered state,
we think that we must take account the spin fluctuation,
because the mean-field approximation overestimates
the magnetic order.

\section*{Acknowledgements}
We would like to thank Kayanuma, Uozumi and Noba for useful discussions.
We also thank the colleague of the quantum physics group of Osaka Prefecture
University.



\end{document}